\documentclass[12pt]{article}
\input{psfig.sty}
\begin{document}
\begin{center}

\vbox{\vspace{20ex}}

{\Large \bf Internal Space-time Symmetries of Massive and Massless
      Particles and their Unification\footnote{presented at the
      International Conference on Supersymmetry and Quantum Field
      Theory, in commemoration of Dmitri Volkov's 75th Birthday,
      Kharkov, Ukraine, 2000.}} \\[6ex]

Y. S. Kim\footnote{electronic address: yskim@physics.umd.edu}\\
{\it Department of Physics, University of Maryland, \\
College Park, Maryland 20742, U.S.A.}
\end{center}

\vspace{4ex}

\begin{abstract}
It is noted that the internal space-time symmetries of relativistic
particles are dictated by Wigner's little groups.  The symmetry of
massive particles is like the three-dimensional rotation group, while
the symmetry of massless particles is locally isomorphic to the
two-dimensional Euclidean group.  It is noted also that, while the
rotational degree of freedom for a massless particle leads to its
helicity, the two translational degrees of freedom correspond to its
gauge degrees of freedom.  It is shown that the $E(2)$-like symmetry of
of massless particles can be obtained as an infinite-momentum and/or
zero-mass limit of the $O(3)$-like symmetry of massive particles.
This mechanism is illustrated in terms of a sphere elongating into
a cylinder.  In this way, the helicity degree of freedom remains
invariant under the Lorentz boost, but the transverse rotational degrees
of freedom become contracted into the gauge degree of freedom.

\end{abstract}

\newpage

\section{Introduction}\label{intro}

If the momentum of a particle is much smaller than its mass, the
energy-momentum relation is $E = p^{2}/2m + mc^{2}$.  If the momentum
is much larger than the mass, the relation is $E = cp$.  These two
different relations can be combined into one covariant formula
$E = \sqrt{m^{2} + p^{2}}$.  This aspect of Einstein's $E = mc^{2}$
is also well known.

In addition, particles have internal space-time variables.  Massive
particles have spins while massless particles have their helicities
and gauge degrees of freedom.
As a ``further content'' of Einstein's $E = mc^{2}$, we shall discuss
that the internal space-time structures of massive and massless particles
can be unified into one covariant package, as $E = \sqrt{m^{2} + p^{2}}$
does for the energy-momentum relation.

The mathematical framework of
this program was developed by Eugene Wigner in 1939~\cite{wig39}.
He constructed the maximal subgroups of the Lorentz group whose
transformations will leave the four-momentum of a given particle
invariant.  These groups are known as Wigner's little groups.
Thus, the transformations of the little groups change the internal
space-time variables of the particle, while leaving its four-momentum
invariant.  The little group is a covariant
entity and takes different forms for the particles moving with
different speeds.

In order to achieve the zero-mass and/or infinite-momentum limit of
the $O(3)$-like little group to obtain the $E(2)$-like little group,
we use the group contraction technique introduced by Inonu and
Wigner~\cite{inonu53}, who obtained the $E(2)$ group by taking
a flat-surface approximation of a spherical surface at the
north pole.  In 1987, Kim and Wigner~\cite{kiwi87jm} observed that
it is also possible to make a cylindrical approximation of the
spherical surface around the equatorial belt.
While the correspondence between $O(3)$ and the $O(3)$-like
little group is transparent, the $E(2)$-like little group contains
both the $E(2)$ group and the cylindrical group~\cite{kiwi90jm}.
We study this aspect in detail in this report.

The space-time symmetries we are discussing in this report are
applicable to all theoretical models of elementary and composite
particles.  Thus, model builders should be aware that their models
should satisfy the these basic symmetries.  They are not going to
build theoretical models which will violate the conservation of
energy, nor are they going to come up with models which will violate
these basic space-time symmetries.

In Sec.~\ref{littleg}, we present a brief history of applications of
the little groups to internal space-time symmetries of relativistic
particles.  It is pointed out in Sec.~\ref{gauge} that the translation-like
transformations of the $E(2)$-like little group corresponds to
gauge transformations.

In Sec.~\ref{o3e2}, we
discuss the contraction of the three-dimensional rotation group to the
two-dimensional Euclidean group.  In Sec.~\ref{contrac}, we
discuss the little group for a massless particle as the
infinite-momentum and/or zero-mass limit of the little group for a
massive particle.

The Lorentz covariance is one of the fundamental issues in modern
physics.  In this paper, we study in this paper for spin-1 particles as
a four-by-four representations of the Lorentz group.  However, there are
many other interesting particles.  Of immediate interest is whether this
formalism can be applied to spin-1/2 particles.  Another interesting
case is a relativistic extended hadrons.  We summarize the current status
of these research lines in Sec.~\ref{further}

\section{Little Groups of the Poincar\'e Group}\label{littleg}

The Poincar\'e group is the group of inhomogeneous Lorentz
transformations, namely Lorentz transformations preceded or followed
by space-time translations.  In order to study this group, we have to
understand first the group of Lorentz transformations, the group of
translations, and how these two groups are combined to form the
Poincar\'e group.  The Poincar\'e group is a semi-direct product of
the Lorentz and translation groups.  The two Casimir operators of
this group correspond to the (mass)$^{2}$ and (spin)$^{2}$ of a given
particle.  Indeed, the particle mass and its spin magnitude are
Lorentz-invariant quantities.

The question then is how to
construct the representations of the Lorentz group which are relevant to
physics.  For this purpose, Wigner in 1939 studied the subgroups of the
Lorentz group whose transformations leave the four-momentum of a given free
particle \cite{wig39}.  The maximal subgroup of the Lorentz group
which leaves the four-momentum invariant is called the little group.
Since the little group leaves the four-momentum invariant, it governs the
internal space-time symmetries of relativistic particles.  Wigner shows in
his paper that the internal space-time symmetries of massive and massless
particles are dictated by the $O(3)$-like and $E(2)$-like little groups
respectively.

The $O(3)$-like little group is locally isomorphic to the three-dimensional
rotation group, which is very familiar to us.  For instance, the group
$SU(2)$ for the electron spin is an $O(3)$-like little group.  The group
$E(2)$ is the Euclidean group in a two-dimensional space, consisting
of translations and rotations on a flat surface.  We are performing
these transformations everyday on ourselves when we move from home to
school.  The mathematics of these Euclidean transformations are also
simple.  However, the group of these transformations are not well
known to us.  In Sec.~\ref{o3e2}, we give a matrix representation
of the $E(2)$ group.

The group of Lorentz transformations consists of three boosts and
three rotations.  The rotations therefore constitute a subgroup of
the Lorentz group.  If a massive particle is at rest, its four-momentum
is invariant under rotations.  Thus the little group for a massive
particle at rest is the three-dimensional rotation group.  Then what is
affected by the rotation?  The answer to this question is very simple.
The particle in general has its spin.  The spin orientation is going
to be affected by the rotation!

If the rest-particle is boosted along the $z$ direction, it will pick
up a non-zero momentum component.  The generators of the $O(3)$ group
will then be boosted.  The boost will take the form of conjugation by
the boost operator.  This boost will not change the Lie algebra of the
rotation group, and the boosted little group will still leave the
boosted four-momentum invariant.  We call this the $O(3)$-like little
group.  If we use the four-vector coordinate $(x, y, z, t)$, the
four-momentum vector for the particle at rest is $(0, 0, 0, m)$, and
the three-dimensional rotation group leaves this four-momentum invariant.
This little group is generated by
\begin{equation}\label{j3}
J_{1} = \pmatrix{0&0&0&0\cr0&0&-i&0\cr0&i&0&0\cr0&0&0&0} , \quad
J_{2} = \pmatrix{0&0&i&0\cr0&0&0&0\cr-i&0&0&0\cr0&0&0&0} , \quad
J_{3} = \pmatrix{0 & -i & 0 & 0 \cr i & 0 & 0 & 0
\cr 0 & 0 & 0 & 0 \cr 0 & 0 & 0 & 0} ,
\end{equation}
which satisfy the commutation relations:
\begin{equation}
 [J_{i}, J_{j}] = i\epsilon_{ijk} J_{k} .
\end{equation}

It is not possible to bring a massless particle to its rest frame.
In his 1939 paper~\cite{wig39}, Wigner observed that the little group
for a massless particle moving along the $z$ axis is generated by the
rotation generator around the $z$ axis, namely $J_{3}$ of Eq.(\ref{j3}),
and two other generators which take the form
\begin{equation}\label{n1n2}
N_{1} = \pmatrix{0 & 0 & -i & i \cr 0 & 0 & 0 & 0
\cr i & 0 & 0 & 0 \cr i & 0 & 0 & 0} ,  \qquad
N_{2} = \pmatrix{0 & 0 & 0 & 0 \cr 0 & 0 & -i & i
\cr 0 & i & 0 & 0 \cr 0 & i & 0 & 0} .
\end{equation}
If we use $K_{i}$ for the boost generator along the i-th axis, these
matrices can be written as
\begin{equation}
N_{1} = K_{1} - J_{2} , \quad N_{2} = K_{2} + J_{1} ,
\end{equation}
with
\begin{equation}
K_{1} = \pmatrix{0&0&0&i\cr0&0&0&0\cr0&0&0&0\cr i&0&0&0} ,  \qquad
K_{2} = \pmatrix{0&0&0&0\cr0&0&0&i\cr0&0&0&0\cr0&i&0&0} .
\end{equation}
The generators $J_{3}, N_{1}$ and $N_{2}$ satisfy the following set
of commutation relations.
\begin{equation}\label{e2lcom}
[N_{1}, N_{2}] = 0 , \quad [J_{3}, N_{1}] = iN_{2} , \quad
[J_{3}, N_{2}] = -iN_{1} .
\end{equation}
In Sec.~\ref{o3e2}, we discuss the generators of the $E(2)$ group.
They are $J_{3}$ which generates rotations around the $z$ axis, and
$P_{1}$ and $P_{2}$ which generate translations along the $x$ and $y$
directions respectively.  If we replace $N_{1}$ and $N_{2}$ by $P_{1}$
and $P_{2}$, the above set of commutation relations becomes the set
given for the $E(2)$ group given in Eq.(\ref{e2com}).  This is the
reason why we say the little group for massless particles is
$E(2)$-like.  Very clearly, the matrices $N_{1}$ and $N_{2}$ generate
Lorentz transformations.

It is not difficult to associate the rotation generator $J_{3}$ with
the helicity degree of freedom of the massless particle.   Then what
physical variable is associated with the $N_{1}$ and $N_{2}$ generators?
Indeed, Wigner was the one who discovered the existence of these
generators, but did not give any physical interpretation to these
translation-like generators.  For this reason, for many years, only
those representations with the zero-eigenvalues of the $N$ operators
were thought to be physically meaningful representations~\cite{wein64}.
It was not until 1971 when Janner and Janssen reported that the
transformations generated by these operators are gauge
transformations~\cite{janner71,kuper76,kim97poz}.  The role of this
translation-like transformation has also been studied for spin-1/2
particles, and it was concluded that the polarization of neutrinos
is due to gauge invariance~\cite{hks82,kim97min}.

Another important development along this line of research is the
application of group contractions to the unifications of the two
different little groups for massive and massless particles.
We always associate the three-dimensional rotation group with a spherical
surface.  Let us consider a circular area of radius 1 kilometer centered
on the north pole of the earth.  Since the radius of the earth is more
than 6,450 times longer, the circular region appears flat.  Thus, within
this region, we use the $E(2)$ symmetry group for this region.  The
validity of this approximation depends on the ratio of the two radii.

\begin{figure}[thb]  
\centerline{\psfig{figure=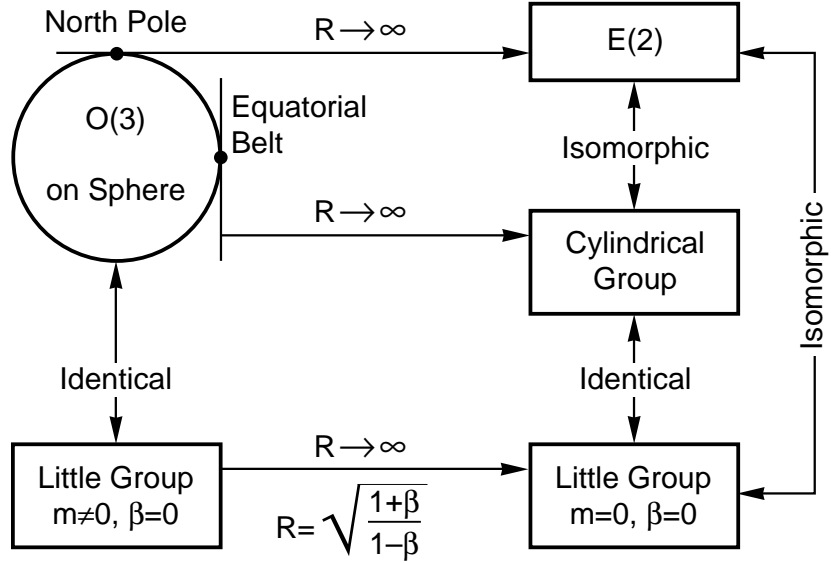,angle=0,height=80mm}}
\caption{Contraction of O(3) to E(2) and to the cylindrical group,
and contraction of the O(3)-like little group to the E(2)-like
little group.  The correspondence between E(2) and the E(2)-like
little group is isomorphic but not identical.  The cylindrical group
is identical to the E(2)-like little group.  The Lorentz boost of
the O(3)-like little group for a massive particle is the same as
the contraction of O(3) to the cylindrical group.} \label{isomor}
\end{figure}

In 1953, Inonu and Wigner formulated this problem as the contraction of
$O(3)$ to $E(2)$~\cite{inonu53}.  How about then the little groups which
are isomorphic to $O(3)$ and $E(2)$?  It is reasonable to expect that the
$E(2)$-like little group be obtained as a limiting case for of the
$O(3)$-like little group for massless particles.  In 1981, it was
observed by Ferrara and Savoy that this limiting process is the Lorentz
boost \cite{ferrara81}.  In 1983, using the
same limiting process as that of Ferrara and Savoy, Han {\it et al}
showed that transverse rotation generators become the generators of
gauge transformations in the limit of infinite momentum and/or zero mass
\cite{hks83pl}.  In 1987, Kim and Wigner showed that the little group for
massless particles is the cylindrical group which is isomorphic to the
$E(2)$ group~\cite{kiwi87jm}.  This is illustrated in Fig.~\ref{isomor}.

\section{Translations and Gauge Transformations}\label{gauge}
It is possible to get the hint
that the $N$ operators generate gauge transformations from Weinberg's
1964 papers~\cite{wein64,hks82}.  But it was not until 1971 when
Janner and Janssen explicitly demonstrated that they generate gauge
transformations~\cite{janner71,kuper76}.  In order to fully appreciate
their work, let us compute the transformation matrix
\begin{equation}
 \exp{(-i(uN_{1} + vN_{2})}
\end{equation}
generated by $N_{1}$ and $N_{2}$.  Then the four-by-four matrix takes
the form
\begin{equation}\label{trans}
\pmatrix{1 & 0 &-u & u \cr 0 & 1 & -v & v \cr
u & v & 1 - (u^{2} + v^{2})/2 & (u^{2} + v^{2})/2 \cr
u & 0 & -(u^{2} + v^{2})/2 & 1 + (u^{2} + v^{2})/2} .
\end{equation}
If we apply this matrix to the four-vector to the four-momentum vector
\begin{equation}\label{4mom}
 p = (0, 0, \omega, \omega)
\end{equation}
of a massless particle, the momentum remains invariant.  It therefore
satisfies the condition for the little group.  If we apply this matrix
to the electromagnetic four-potential
\begin{equation}
 A = (A_{1}, A_{2}, A_{3}, A_{0}) \exp{(i(kz -\omega t))} ,
\end{equation}
with $A_{3} = A_{0}$ which is the Lorentz condition,
the result is a gauge transformation.  This is what Janner and Janssen
discovered in their 1971 and 1972 papers~\cite{janner71}. Thus
the matrices $N_{1}$ and $N_{2}$ generate gauge transformations.

\section{Contraction of O(3) to E(2)}\label{o3e2}
In this Appendix, we explain what the $E(2)$ group is.  We then
explain how we can obtain this group from the three-dimensional
rotation group by making a flat-surface or cylindrical approximation.
This contraction procedure will give a clue to obtaining the $E(2)$-like
symmetry for massless particles from the $O(3)$-like symmetry for
massive particles by making the infinite-momentum limit.

The $E(2)$ transformations consist of rotation and two translations on
a flat plane.  Let us start with the  rotation matrix applicable to
the column vector $(x, y, 1)$:
\begin{equation}\label{rot}
 R(\theta) = \pmatrix{\cos\theta & -\sin\theta & 0 \cr
\sin\theta & \cos\theta & 0 \cr 0 & 0 & 1} .
\end{equation}
Let us then consider the translation matrix:
\begin{equation}
 T(a, b) = \pmatrix{1 & 0 & a \cr 0 & 1 & b \cr 0 & 0 & 1} .
\end{equation}
If we take the product $T(a, b) R(\theta)$,
\begin{equation}\label{eucl}
E(a, b, \theta) = T(a, b) R(\theta)  
= \pmatrix{\cos\theta & -\sin\theta & a \cr
\sin\theta & \cos\theta & b \cr 0 & 0 & 1} .
\end{equation}
This is the Euclidean transformation matrix applicable to the
two-dimensional $x y$ plane.  The matrices $R(\theta)$ and $T(a,b)$
represent the rotation and translation subgroups respectively.  The
above expression is not a direct product because $R(\theta)$ does not
commute with $T(a,b)$.  The translations constitute an Abelian invariant
subgroup because two different $T$ matrices commute with each other,
and because
\begin{equation}
 R(\theta) T(a,b) R^{-1}(\theta) = T(a',b') .
\end{equation}
The rotation subgroup is not invariant because the conjugation
\begin{equation}
 T(a,b) R(\theta) T^{-1}(a,b)
\end{equation}
does not lead to another rotation.

We can write the above transformation matrix in terms of generators.
The rotation is generated by
\begin{equation}
 J_{3} = \pmatrix{0 & -i & 0 \cr i & 0 & 0 \cr 0 & 0 & 0} .
\end{equation}
The translations are generated by
\begin{equation}
 P_{1} = \pmatrix{0 & 0 & i \cr 0 & 0 & 0 \cr 0 & 0 & 0} , \quad
P_{2} = \pmatrix{0 & 0 & 0 \cr 0 & 0 & i \cr 0 & 0 & 0} .
\end{equation}
These generators satisfy the commutation relations:
\begin{equation}\label{e2com}
[P_{1}, P_{2}] = 0 , \qquad [J_{3}, P_{1}] = iP_{2} , \qquad
[J_{3}, P_{2}] = -iP_{1} .
\end{equation}
This $E(2)$ group is not only convenient for illustrating the groups
containing an Abelian invariant subgroup, but also occupies an
important place in constructing representations for the little
group for massless particles, since the little group for massless
particles is locally isomorphic to the above $E(2)$ group.

The contraction of $O(3)$ to $E(2)$ is well known and is often called
the Inonu-Wigner contraction~\cite{inonu53}.  The question is whether
the $E(2)$-like little group can be obtained from the $O(3)$-like
little group.  In order to answer this question, let us closely look
at the original form of the Inonu-Wigner contraction.  We start with
the generators of $O(3)$.  The $J_{3}$ matrix is given in Eq.(\ref{j3}),
and
\begin{equation}\label{o3gen}
J_{2} = \pmatrix{0&0&i\cr0&0&0\cr-i&0&0} ,  \qquad
J_{3} = \pmatrix{0&-i&0\cr i &0&0\cr0&0&0} .
\end{equation}
The Euclidean group $E(2)$ is generated by $J_{3}, P_{1}$ and $P_{2}$,
and their Lie algebra has been discussed in Sec.~\ref{intro}.

\begin{figure}[thb]  
\centerline{\psfig{figure=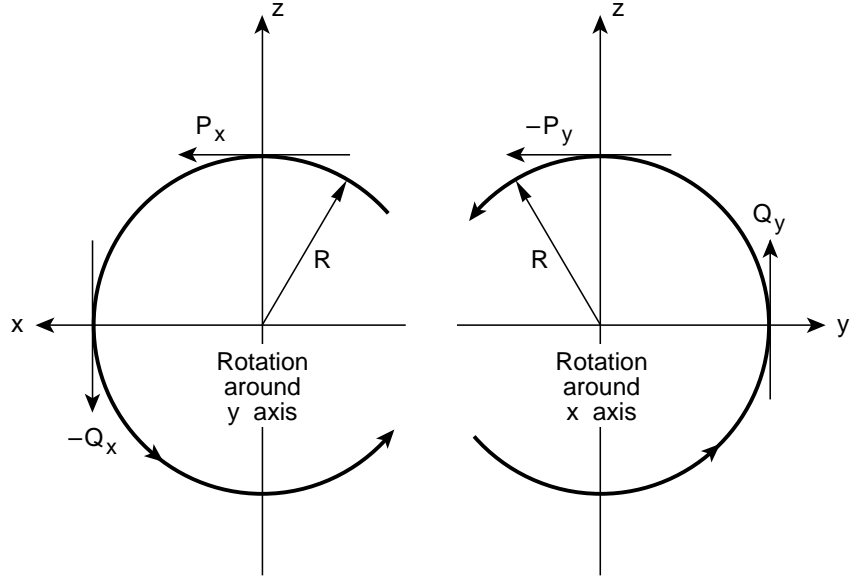,angle=0,height=80mm}}
\caption{North-pole and Equatorial-belt approximations.  The north-pole
approximation leads to the contraction of O(3) to E(2).  The
equatorial-belt approximation leads corresponds to the contraction the
cylindrical group.}\label{equator}
\end{figure}

Let us transpose the Lie algebra of the $E(2)$ group.  Then $P_{1}$ and
$P_{2}$ become $Q_{1}$ and $Q_{2}$ respectively, where
\begin{equation}
Q_{1} = \pmatrix{0&0&0\cr0&0&0\cr i &0&0} , \qquad
Q_{2} = \pmatrix{0&0&0\cr0&0&0\cr0&i&0} .
\end{equation}
Together with $J_{3}$, these generators satisfy the
same set of commutation relations as that for
$J_{3}, P_{1}$, and $P_{2}$ given in Eq.(\ref{e2com}):
\begin{equation}
[Q_{1}, Q_{2}] = 0 , \qquad [J_{3}, Q_{1}] = iQ_{2} , \qquad
[J_{3}, Q_{2}] = -iQ_{1} .
\end{equation}
These matrices generate transformations of a point on a circular
cylinder.  Rotations around the cylindrical axis are generated by
$J_{3}$.  The matrices $Q_{1}$ and $Q_{2}$ generate translations along
the direction of $z$ axis.  The group generated by these three matrices
is called the {\it cylindrical group}~\cite{kiwi87jm,kiwi90jm}.

We can achieve the contractions to the Euclidean and cylindrical groups
by taking the large-radius limits of
\begin{equation}\label{inonucont}
 P_{1} = {1\over R} B^{-1} J_{2} B ,
\qquad P_{2} = -{1\over R} B^{-1} J_{1} B ,
\end{equation}
and
\begin{equation}
 Q_{1} = -{1\over R}B J_{2}B^{-1} , \qquad
Q_{2} = {1\over R} B J_{1} B^{-1} ,
\end{equation}
where
\begin{equation}\label{bmatrix}
 B(R) = \pmatrix{1&0&0\cr0&1&0\cr0&0&R}  .
\end{equation}
The vector spaces to which the above generators are applicable are
$(x, y, z/R)$ and $(x, y, Rz)$ for the Euclidean and cylindrical groups
respectively.  They can be regarded as the north-pole and equatorial-belt
approximations of the spherical surface respectively~\cite{kiwi87jm}.
Fig.~\ref{equator} illustrates how the Euclidean and cylindrical
contractions are made.

\section{Contraction of O(3)-like Little Group to E(2)-like Little
Group}\label{contrac}

Since $P_{1} (P_{2})$ commutes with $Q_{2} (Q_{1})$, we can consider the
following combination of generators.
\begin{equation}
 F_{1} = P_{1} + Q_{1} , \quad F_{2} = P_{2} + Q_{2} .
\end{equation}
Then these operators also satisfy the commutation relations:
\begin{equation}\label{commuf}
[F_{1}, F_{2}] = 0 , \qquad [J_{3}, F_{1}] = iF_{2} ,  \qquad
[J_{3}, F_{2}] = -iF_{1} .
\end{equation}
However, we cannot make this addition using the three-by-three matrices
for $P_{i}$ and $Q_{i}$ to construct three-by-three matrices for $F_{1}$
and $F_{2}$, because the vector spaces are different for the $P_{i}$ and
$Q_{i}$ representations.  We can accommodate this difference by creating
two different $z$ coordinates, one with a contracted $z$ and the other
with an expanded $z$, namely $(x, y, Rz, z/R)$.  Then the generators
become
\begin{equation}
P_{1} = \pmatrix{0&0&0&i\cr0&0&0&0\cr0&0&0&0\cr0&0&0&0} ,  \qquad
P_{2} = \pmatrix{0&0&0&0\cr0&0&0&i\cr0&0&0&0\cr0&0&0&0} ,
\end{equation}
and
\begin{equation}
Q_{1} = \pmatrix{0&0&0&0\cr0&0&0&0\cr i &0&0&0\cr0&0&0&0} ,  \qquad
Q_{2} = \pmatrix{0&0&0&0\cr0&0&0&0\cr0&i&0&0\cr0&0&0&0} .
\end{equation}
Then $F_{1}$ and $F_{2}$ will take the form
\begin{equation}\label{f1f2}
F_{1} = \pmatrix{0&0&0&i\cr0&0&0&0\cr i &0&0&0\cr0&0&0&0} ,  \qquad
F_{2} = \pmatrix{0&0&0&0\cr0&0&0&i\cr0&i&0&0\cr0&0&0&0} .
\end{equation}
The rotation generator $J_{3}$ takes the form of Eq.(\ref{j3}).
These four-by-four matrices satisfy the E(2)-like commutation relations
of Eq.(\ref{commuf}).

\begin{figure}[thb]  
\centerline{\psfig{figure=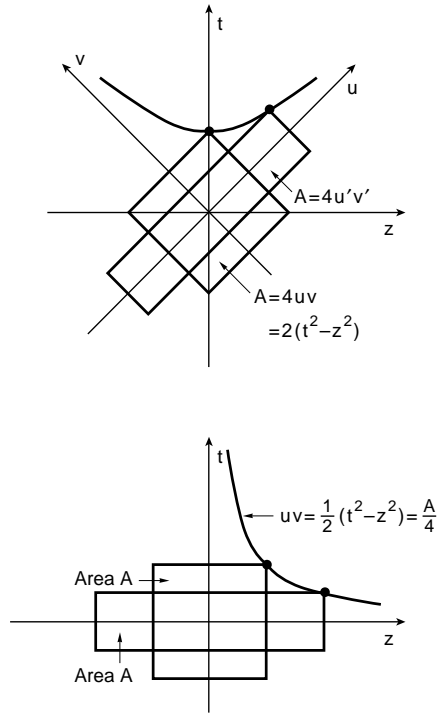,angle=0,height=95mm}}
\caption{Light-cone coordinates.  When the system is Lorentz-boosted,
one of the axes expands while the other becomes contracted.  Both
the expansion and contraction are needed for the contraction of
the $O(3)$-like little group to $E(2)$-like little group.}\label{lcone}
\end{figure}

Now the $B$ matrix of Eq.(\ref{bmatrix}), can be expanded to
\begin{equation}\label{bmatrix2}
 B(R) = \pmatrix{1&0&0&0\cr0&1&0&0\cr0&0&R&0\cr0&0&0&1/R} .
\end{equation}
This matrix includes both the contraction and expansion in the light-cone
coordinate system, as illustrated in Fig.~\ref{lcone}.
If we make a similarity transformation on the above form using the matrix
\begin{equation}\label{simil}
 \pmatrix{1&0&0&0\cr0&1&0&0\cr0&0&1/\sqrt{2} &-1/\sqrt{2}
\cr0&0&1/\sqrt{2}&1/\sqrt{2}} ,
\end{equation}
which performs a 45-degree rotation of the third and fourth coordinates,
then this matrix becomes
\begin{equation}\label{simil2}
 \pmatrix{1&0&0&0\cr0&1&0&0\cr0&0 & \cosh\eta & \sinh\eta
\cr0 & 0 & \sinh\eta & \cosh\eta} ,
\end{equation}
with $R = e^\eta$.  This form is the Lorentz boost matrix along the $z$
direction.  If we start with the set of expanded rotation generators
$J_{3}$ of Eq.(\ref{j3}), and
perform the same operation as the original Inonu-Wigner contraction
given in Eq.(\ref{inonucont}), the result is
\begin{equation}
 N_{1} = {1\over R} B^{-1} J_{2} B ,
\quad N_{2} = -{1\over R} B^{-1} J_{1} B ,
\end{equation}
where $N_{1}$ and $N_{2}$ are given in Eq.(\ref{n1n2}).  The generators
$N_{1}$ and $N_{2}$ are the contracted $J_{2}$ and $J_{1}$ respectively
in the infinite-momentum and/or zero-mass limit.

It was noted in Sec.~\ref{gauge} that $N_{1}$ and $N_{2}$ generate
gauge transformations on massless particles.  Thus the contraction of
the transverse rotations leads to gauge transformations.

\section{Further Considerations}\label{further}
We have seen in this report that Wigner's $O(3)$-like little group can
be contracted into the $E(2)$-like little group for massless particles.
Here, we worked out explicitly for the spin-1 case, but this mechanism
should be applicable to all other spins.  Of particular interest is
spin-1/2 particles.  This has been studied by Han, Kim and Son~\cite{hks82}.
They noted that there are also gauge transformations for spin-1/2
particles, and the polarization of neutrinos is a consequence of gauge
invariance.  It has also been shown that the gauge dependence of spin-1
particles can be traced to the gauge variable of the spin-1/2
particle~\cite{hks86jm}.  It would be very interesting to see how
the present formalism is applicable to higher-spin particles.

Another case of interest is the space-time symmetry of relativistic
extended particles.  In 1973~\cite{kn73}, Kim and Noz constructed
a ground-state harmonic oscillator wave function which can be
Lorentz boosted.  It was later found that this oscillator formalisms
can be extended to represent the $O(3)$-like little
group~\cite{kno79,knp86}.  This oscillator formalism has a stormy
history because it ultimately plays a pivotal role in combining
quantum mechanics and special relativity~\cite{dir45,yuka53}.

With these wave functions, we propose to solve the following problem in
high-energy physics.  The quark model works well when hadrons are at
rest or move slowly.  However, when they move with speed close to that
of light, they appear as a collection of infinite-number of
partons~\cite{fey69}.  The question then is whether the parton model is
a Lorentz-boosted quark model.  This question has been addressed
before~\cite{kn77par,kim89}, but it can generate more interesting
problems~\cite{kiwi90pl}.  The present situation is presented in the
following table.

\begin{table}[thb]

\caption{Massive and massless particles in one package.  Wigner's
little group unifies the internal space-time symmetries for massive and
massless particles.  It is a great challenge for us to find
another unification: the unification of the quark and parton pictures in
high-energy physics.}

\vspace{3mm}

\begin{center}

\begin{tabular}{lccc}

\hline
{}&{}&{}&{}\\
{} & Massive, Slow \hspace{6mm} & COVARIANCE \hspace{6mm}&
Massless, Fast \\[4mm]\hline
{}&{}&{}&{}\\
Energy- & {}  & Einstein's & {} \\
Momentum & $E = p^{2}/2m$ & $ E = [p^{2} + m^{2}]^{1/2}$ & $E = cp$
\\[4mm]\hline
{}&{}&{}&{}\\
Internal & $S_{3}$ & {}  &  $S_{3}$ \\[-1mm]
Space-time &{} & Wigner's  & {} \\ [-1mm]
Symmetry & $S_{1}, S_{2}$ & Little Group & Gauge Trans. \\[4mm]\hline
{}&{}&{}&{}\\
Relativistic & {} & One  &  {} \\[-1mm]
Extended & Quark Model & Covariant  & Parton Model\\ [-1mm]
Particles & {} & Theory &{} {} \\[4mm]\hline

\end{tabular}

\end{center}
\end{table}

\section*{Acknowledgments}
The author would like to thank D. Sorokin and other members of the
local organizing committee for inviting him for the
important occasion to commemorate the 75th anniversary of Prof.
Dmitri Volkov's birth.  We were very sorry for not being able to
greet Prof. Volkov, but we all felt his influence throughout the
conference.  Mrs. Volkov was very kind to greet us.  The author is
grateful to the citizens of Kharkov for providing warm hospitality to him.

\end{document}